%% file: rdf-hunter.tex
\documentclass{llncs}

\usepackage{times}
\usepackage{url}
\usepackage{graphicx}
\usepackage{ctable}
\usepackage{verbatim}
\usepackage{listings}
\usepackage{multirow}
\usepackage{caption}
\usepackage{subfigure}
\usepackage{amsmath}
\usepackage{parcolumns}
\usepackage{paralist}
\usepackage{todo}
\usepackage[ruled, vlined, linesnumbered]{algorithm2e}

\newcommand{\andl}{\ensuremath{\ \wedge\ }}   

\newcommand{\sfs}{\scriptsize\sffamily}

\begin{document}
\allowdisplaybreaks
\title{RDF-Hunter: Automatically Crowdsourcing the Execution of Queries Against RDF Data Sets}
\author{Maribel Acosta\inst{1} \and Elena Simperl\inst{2} \and  Fabian Fl\"{o}ck\inst{3} \and\\ Maria-Esther Vidal\inst{4} \and Rudi Studer\inst{1}}
\institute{$^1$ Institute AIFB, Karlsruhe Institute of Technology\\
          \email{\{maribel.acosta,rudi.studer\}@kit.edu}\\
           $^2$Web Science and Internet Research Group, University of Southampton\\
          \email{e.simperl@soton.ac.uk}\\
	  $^3$Computational Social Science Group, GESIS - Leibniz Institute for the Social Sciences\\
          \email{fabian.floeck@gesis.org}\\
          $^4$Computer Science Department, Universidad Sim\'on Bol\'{\i}var \\
          \email{mvidal@ldc.usb.ve}}

\maketitle

\input{sections/abstract}
\input{sections/introduction}
\input{sections/relatedwork}

\input{sections/motivation}
\input{sections/approach}
\input{sections/experiments}

\input{sections/conclusions}

\bibliographystyle{abbrv} 
\bibliography{references}

\end{document}

%% file: sections/abstract.tex
\begin{abstract}
In the last years, a large number of RDF data sets has become available on the Web. However, due to the semi-structured nature of RDF data, missing values affect  answer completeness of queries that are posed against this data.
To overcome this limitation, we propose RDF-Hunter, a novel hybrid query processing approach that brings together machine and human computation to execute queries against RDF data. 
We develop a novel quality model and query engine in order to enable RDF-Hunter to on the fly decide  which parts of a query should be executed  through conventional technology or crowd computing. 
To evaluate RDF-Hunter, we created a collection of 50 SPARQL queries against the DBpedia data set, executed them using our hybrid query engine, and analyzed the accuracy of the outcomes obtained from the crowd. The experiments clearly show that the overall approach is  feasible and produces query results that reliably and significantly enhance completeness of automatic query processing responses.
\end{abstract}

%% file: sections/introduction.tex
\section{Introduction}
\label{sec:introduction}
Linked Open Data (LOD) initiatives have fostered the publication of Linked Data (LD) on almost any subject~\cite{LOD2014}. 
The majority of these data artifacts have been created by integrating multiple, typically heterogenous sources, and contain a fair share of missing values. Yet, due to the semi-structured nature of RDF data, such incompleteness cannot be easily detected, with negative effects on query processing. For example, running a query against the DBpedia data set\footnote{SPARQL endpoint: \url{http://dbpedia.org/sparql}} that asks for {\it movies, including their producers, that have been filmed in New York City by Universal Pictures} returns no producers for 14\% of the movies in the result set. With cases like this being a common occurrence in LD applications, further techniques are needed to improve this data aspect and subsequent query processing results.
Recent research suggests that microtask crowdsourcing provides a platform for implementing effective hybrid approaches to Linked Data quality assessment~\cite{acosta2013}. The relational data base community has embraced similar ideas to design advanced query processing systems that combine human and computational intelligence ~\cite{DBLP:journals/pvldb/0002KMMO12,deco2012vldb,DBLP:journals/pvldb/ParkW13,DBLP:conf/icde/TrushkowskyKFS13}. However, 
 most of the existing proposals focus on manually specifying those parts of the query that should resort to human involvement, typically devising bespoke query languages and extensions that would be applied manually on top of established data base technology. Such an approach is less feasible for an LD scenario, which is confronted with decentralized large volumes of semi-structured data.
In this work we present RDF-Hunter, the first system that, by exploiting the characteristics of RDF data, \emph{automatically identifies the exact portions of a query against an RDF data set that should be processed by the crowd in order to augment answer completeness}. Going back to our example, RDF-Hunter will assess that the sub-query asking for movie producers needs to be outsourced to the crowd in order to collect the missing information in the DBpedia data set; it will then autonomously determine how to set up the corresponding crowdsourcing task and execute it against a microtask platform.
In a nutshell, RDF-Hunter is a hybrid query processing system that combines human and computer capabilities to run queries against RDF data sets. Its aim is to enhance the  answer  completeness of SPARQL queries by finding missing values in the data set via microtask crowdsourcing. 
Our solution provides a highly flexible crowdsourcing-enabled SPARQL query execution: no extensions to SPARQL or RDF are required, and the user can configure the level of expected answer completeness in each query execution.
We define a quality model for completeness of RDF data sets. 
RDF-Hunter implements query decomposition techniques able to on the fly decide  the parts of a SPARQL query that are potentially affected by missing data values and should resort to the crowd.  The query engine combines crowd and intermediary automatically computed results. During execution time, RDF-Hunter collects information about the types of queries the crowd is likely to be able to solve accurately. 
 
To evaluate RDF-Hunter, we crafted a collection of 50 SPARQL queries against DBpedia (version 2014), executed them with our system, and analyzed the quality of the crowd  answers. 
The goal of the experiments was to assess the  answer completeness  produced by RDF-Hunter when queries were executed against a SPARQL endpoint and the CrowdFlower\footnote{\url{http://www.crowdflower.com/}}  crowdsourcing platform. The empirical results clearly show that the overall approach is not only feasible  but can reliably augment response completeness.

The contributions of our work can be summarized as follows:
\vspace{-0.2cm}
\begin{itemize}
\item the design of a quality model for estimating RDF completeness;
\item a proposal for interpreting crowd knowledge; 
\item a novel query planner and a query execution engine able to decide which parts of a SPARQL query will be executed against an RDF data set and the crowd;
\item an extensive benchmark composed of 50 SPARQL queries to study and evaluate the answer completeness of query processing systems; and
\item an extensive empirical evaluation using the DBpedia public SPARQL endpoint and the CrowdFlower microtask platform.
\end{itemize}

%% file: sections/relatedwork.tex
\section{Related Work}
\label{sec:relatedwork}
\vspace{-0.1cm}
The database community has proposed several human/computer query processing architectures for relational data. 
Approaches such as CrowdDB~\cite{franklin2011,DBLP:conf/icde/TrushkowskyKFS13}, Deco~\cite{deco2012vldb}, and Qurk~\cite{DBLP:journals/pvldb/0002KMMO12} target scenarios in which existing microtask platforms are directly embedded in query processing systems. 
CrowdDB~\cite{franklin2011,DBLP:conf/icde/TrushkowskyKFS13} provides SQL-like data definition and query languages to support hybrid query execution, and attempts to reduce the number of tasks to be outsourced by exploiting structural properties of the relational data~\cite{DBLP:conf/icde/TrushkowskyKFS13}. 
Deco~\cite{deco2012vldb} implements caching strategies to reuse previously crowdsourced data. Additionally, Deco~\cite{DBLP:journals/pvldb/ParkW13} and Qurk~\cite{DBLP:journals/pvldb/0002KMMO12}  provide a set of physical operators 
and  models to estimate selectivities and cardinalities. These statistics  in conjunction with the physical operators allow to define physical plans that reduce execution time, monetary cost,  and  number of tasks. 
By contrast, we propose a {\it quality model} that not only exploits the structure of RDF data sets, but also values of {\it disagreement} and {\it uncertainty} of the  crowdsourced data. Additionally, we  devise a {\it query planner} that relies on the quality model  and performs query decomposition techniques able to {\it automatically} generate data-informed processing pipelines that use crowd intelligence effectively. In this way, the query planner makes sure that human contribution is sought only in those cases in which it will most likely lead to result improvements. This speeds up the overall query execution, and reduces both the costs and the average time that is needed to obtain the crowdsourced answers. Further, RDF-Hunter liberates users from the task of manually selecting the parts of the queries that will be evaluated by the crowd and the ones that will be posed against the RDF engines. 


Additionally, crowdsourcing has also shown to be feasible for other scenarios related to Semantic Web technologies. 
Amsterdamer et al. propose OASSIS~\cite{DBLP:conf/sigmod/AmsterdamerDMNS14}, 
 a query-driven crowdsourcing platform that responds to user information needs. OASSIS combines  general knowledge from ontologies with frequent patterns mined from  personal  data collected from the crowd.
OASSIS provides a SPARQL-like query language where users specify  sub-queries that will be evaluated against the ontology and the ones that will be mined from the crowd.  Additionally, the OASSIS query engine is able to order the execution of the sub-queries in way that questions posed to the crowd are minimized.
LODRefine,\footnote{\url{http://code.zemanta.com/sparkica/}} an LD integration tool, has made available an extension  that allows to  manually configure and run specific data matchmaking tasks on CrowdFlower.
 Although these approaches address different LD management problems, they require the user intervention on the definition of the crowd-based workflows that will be evaluated to solve the corresponding LD management problem. 
In contrast, RDF-Hunter  automatically creates hybrid query-driven workflows, and combines the results obtained from both RDF data sets and the crowd to enhance  completeness during query processing.

%% file: sections/motivation.tex
\section{Motivating Example}
\label{sec:motivation}
Consider the  SPARQL query in Listing~\ref{motivatingquery1} to be issued against the DBpedia endpoint. 
This query retrieves information about capitals in Europe and their respective country. 
When executing the query, the total number of answers is $47$. 
This means that DBpedia contains $47$ entities that are classified as European capitals (triple pattern $4$) {\it and} that are linked to their corresponding country (triple pattern $5$). 
However, by executing only triple pattern $4$, it is revealed that DBpedia contains $56$ bindings for European capitals. 
This suggests that the completeness of this portion of the data set is 0.84.  
\begin{lstlisting}[basicstyle=\scriptsize\sffamily, language=SPARQL, caption={SPARQL query against DBpedia to select cities and countries such that the cities are capitals in Europe}, label=motivatingquery1, numbers=left, numberstyle=\scriptsize]
PREFIX dbpedia-yago: <http://dbpedia.org/class/yago/>
PREFIX dbpedia-owl: <http://dbpedia.org/ontology/>
SELECT DISTINCT ?city ?country WHERE {
  ?city a dbpedia-yago:CapitalsInEurope .
  ?city dbpedia-owl:country ?country .}
\end{lstlisting}
%

We crowdsourced the missing values of the previous SPARQL query via microtasks submitted to CrowdFlower. 
Table~\ref{table:example_results1} reports on: (i) the results obtained from the crowd (the value for the variable {\sfs ?country}); (ii) the crowd's confidence, denoted as $\gamma \in [0.0,1.0]$, provided by the platform; and (iii) $\phi$, the normalized average of the familiarity of the crowd workers with the topic on a scale from 1 to 7 (which we inquired for). 
For the query from Listing~\ref{motivatingquery1}, the crowd answered that $8$ out of the $9$ cities are located in a country.  
The crowd submitted answers with high values of confidence, on average 0.89.  
61\%  of the participants in these tasks claimed that they were familiar with the topic. 

%
\begin{table}[b!]
\centering
\caption{Crowdsourcing results for query from Listing~\ref{motivatingquery1}. $\gamma \in [0.0,1.0]$ is the crowd's confidence, and $\phi \in [0.0,1.0]$  is the average of the crowd's familiarity to the topic. The {\sfs db} prefix corresponds to {\sfs $<$http://dbpedia.org/resource/$>$}} 
\label{table:example_results1}
{\scriptsize
\begin{tabular}{|l|l|l|l|}
\hline
{\bf Crowdsourced} & {\bf Crowd's answers} & {\bf $\gamma$ } & {\bf $\phi$}\\
{\bf instances of ?city } & {\bf for ?country} & & \\
\hline
{\sf db:Chi\c{s}in\u{a}u} & {\sf db:Moldova}& 0.833&0.476\\
{\sf db:Edinburgh} & {\sf db:Scotland} & 1.0 &0.761\\
{\sf db:Episkopi\_Cantonment} & {\sf db:Akrotiri\_and\_Dhekelia} & 0.833&0.404\\
{\sf db:Gibraltar}& {\sf db:United\_Kingdom} & 0.666 & 0.69\\
{\sf db:Helsinki} & {\sf db:Finland}& 1.0&0.743\\
{\sf db:Madrid}& {\sf db:Spain} & 1.0&0.976\\
{\sf db:Pristina}& {\sf db:Kosovo} & 1.0&0.714\\
{\sf db:Vatican\_City}& {\sf db:Vatican\_City} & 0.833&0.81\\
\hline
{\sf db:Monaco}& (No value) & 0.80&0.743\\
\hline
\end{tabular}}
\end{table}
The query presented above illustrates the aspects of SPARQL query execution which are likely to benefit most from crowdsourcing: 1) the portion of the RDF data set contains missing values; and 2) the crowd has the skills to complete the missing portions of the data sets. 
These two properties allow to devise effective solutions for crowdsourcing query execution that are able to scale to large data sets.
A na\"{i}ve approach that submits to the crowd every single triple pattern contained in a query is not feasible,  since the amount of data subject to human assessment (and thus the associated cost) would be very large. Therefore, we propose an approach that exploits the structure of RDF data sets and information about the crowd to on the fly decide which parts of the query require human intervention to be able to scale up to the LOD data sets.

%% file: sections/approach.tex
\section{Our Approach}
\label{approach}
\input{sections/approach-overview}

\input{sections/approach-model}

\input{sections/approach-crowd}

\input{sections/approach-microtask}

\input{sections/approach-optimizer}
\input{sections/approach-engine}


%% file: sections/approach-overview.tex
\begin{figure*}[t!]
\begin{center}
\includegraphics[width=0.80\textwidth]{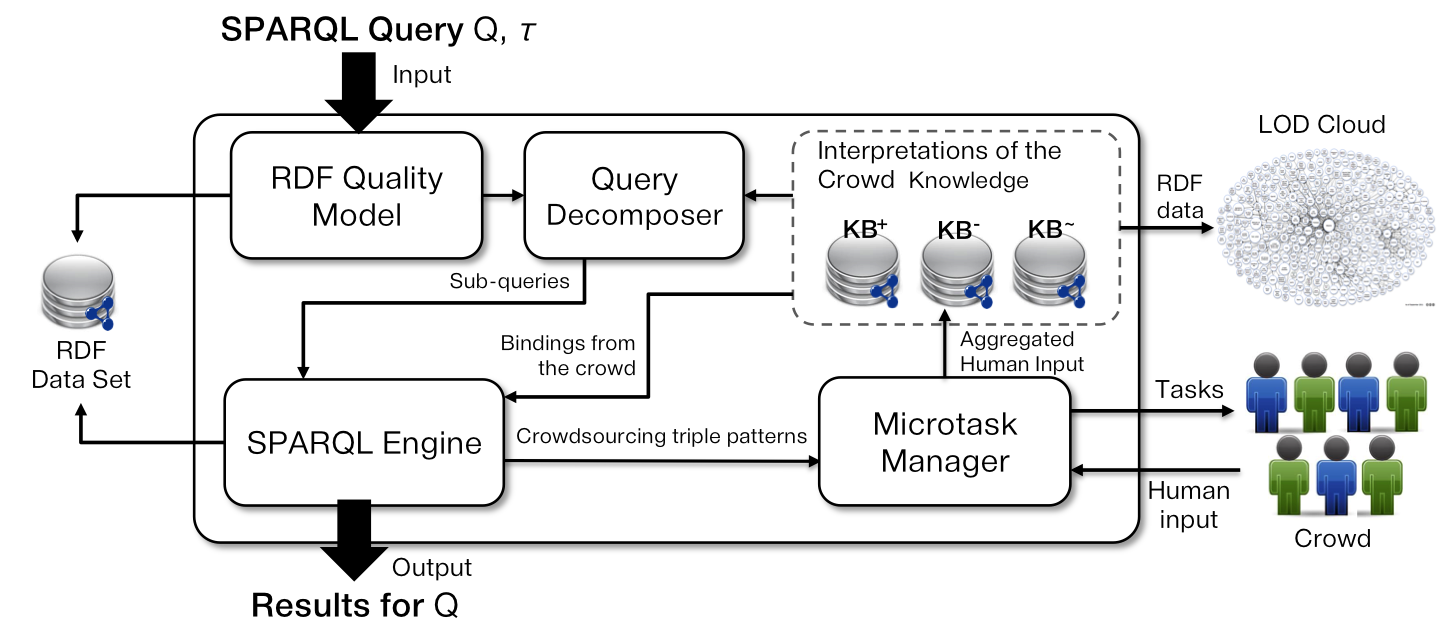}
\caption{The RDF-Hunter architecture}
\label{fig:architecture}
\end{center}
\end{figure*}

Figure~\ref{fig:architecture} depicts the components of RDF-Hunter, which 
receives as input a SPARQL query $Q$ and a quality threshold $\tau$.
The  {\it RDF Quality Model} estimates the completeness of the portions of the data set that yields results for $Q$. 
The {\it Query Decomposer} 
 generates sub-queries from $Q$, taking into consideration $\tau$, the quality model and the human input stored in the {\it Interpretations of the Crowd Knowledge}.  
The sub-queries are executed by the {\it SPARQL Engine}, which contacts the RDF data set and sends the crowdsourced RDF triples to the {\it Microtask Manager}. 
The microtask manager generates the human tasks and submits them to the microtask platform. 
The SPARQL engine combines the results retrieved from the data set with the human input to produce the results for $Q$. 
%

%% file: sections/approach-model.tex
\subsection{RDF Quality Model for Completeness}
\label{sec:qualmodel}
We propose a model to estimate the completeness of RDF data sets.  Our model captures  the multiplicity of predicates at the level of RDF resources  and classes. These multiplicities are used to   
compute the completeness of the RDF resources of the data set with respect to the predicates that characterize these resources. 
First, we define the multiplicity of a resource $s$  with respect to predicate $p$, named $M_D(s|p)$ as the number of objects associated with the resource $s$ through the predicate $p$.

\begin{definition} {\it (Predicate Multiplicity of an RDF Resource).} 
\label{def:model}
Given an RDF resource $s$ occurring in the data set $D$, the multiplicity of the predicate $p$ for the resource $s$ is:
$$M_D(s|p) :=  |\{o | (s,p,o) \in D\}| $$
\end{definition}

\begin{figure}[b!]
\begin{center}
\includegraphics[width=0.95\textwidth]{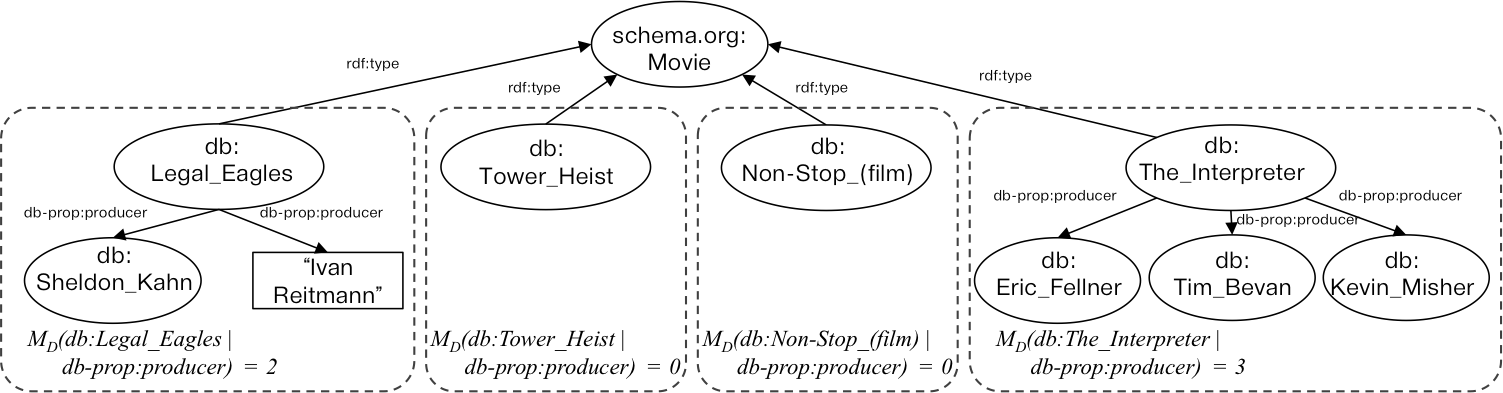}
\vspace{5px}
\caption{Portion of the DBpedia data set for movies}
\label{fig:cms}
\end{center}
\end{figure}

Consider the RDF graph depicted in Figure~\ref{fig:cms}, where ovals represent URIs and rectangles denote literals.  
Edges correspond to relationships between nodes annotated with the corresponding predicate.  This RDF graph contains four  nodes of type {\sfs schema.org:Movie}. In this figure, movies are enclosed with the nodes that represent their producers; multiplicity of the predicate {\sfs db-prop:producer} is presented for each movie. For example, the resource $s= $ {\sfs db:The\_Interpreter} has  three different values for the predicate $p= $ {\sfs db-prop:producer}, therefore, $M_D(s|p)$ is 3 in this case.
%

Next, we define the multiplicity of a class $C$ with respect to a predicate $p$,  named $AM_D(C|p)$, as the aggregated number of objects associated through the predicate $p$ with resources $s$ belonging to the class $C$.  


\begin{definition} {\it (Aggregated Multiplicity of a Class).} 
\label{def:model}
For each class $C$ occurring in the RDF data set $D$, the aggregated multiplicity of $C$ over the predicate $p$ is: 
\begin{equation*}
AM_D(C|p) :=  \lceil ( F(\{M_D(s|p) | (s, p, o) \in D \andl (s, a, C) \in D\})) \rceil
\end{equation*}
\begin{itemize}
\item $(s, a, C)$ corresponds to the triple {\sfs (s, rdf:type, C)}, which means that the subject $s$ belongs to the class  $C$.
\item $F (.)$ is an aggregation function, e.g., the median.
\end{itemize}
\end{definition}

Suppose the class {\sfs schema.org:Movie}  comprises only the four movies in Figure~\ref{fig:cms}, and the median is the aggregation function. The  
multiplicity of {\sfs schema.org:Movie} with respect to the predicate {\sfs df-prop:producer}, i.e., $AM_D$({\sfs schema.org:Movie}$|${\sfs db-prop:producer}), is 3. 
Note that movies {\sfs db:Tower\_Heist} and {\sfs db:Non-Stop\_(film)} are not considered because they are not related to any producer in this data set, and they are not considered in the computation of the aggregation function $F$.

Finally, the completeness of an RDF resource $s$ with respect to a predicate $p$ is defined as: the result of normalizing the multiplicity of $s$ with respect to $p$ 
by the maximum value of multiplicity of the classes to which $s$ belongs. 

\begin{definition}{\it (Completeness of an RDF Resource with Respect to a Predicate.)}
Given an RDF resource $s$ and a predicate $p$ occurring in the data set $D$. Let $C_1$, ..., $C_n$ be the classes in $D$ such that $(s, a, C_1) \in D, ... , (s, a, C_n) \in D$. The completeness of $s$ with respect to $p$ in the data set $D$ is defined as follows: 
\begin{equation*}
\hbox{\it Comp}_D(s|p) := 
\left\{
\begin{array}{lll}
 \frac{M_D(s|p)}{AM_D(C'|p)}   & \mbox{if } AM_D(C'|p) \neq 0 \\
 1  & \mbox{otherwise}
\end{array}
\right.
\end{equation*}
where $AM_D(C'|p) = max(AM_D(C_1|p), ... ,AM_D(C_n|p))$
\end{definition}

Suppose  the movie {\sfs db:The\_Interpreter} also belongs to the class {\sfs http://dbpedia.org/ontology/Film}, and the aggregated multiplicity of this class with respect to  the predicate {\sfs db-prop:producer} is 5.
Then, the completeness of   {\sfs db:The\_Interpreter} is 0.6, indicating that 40\% of the producers of this movie are not represented in this data set.

%% file: sections/approach-crowd.tex
\subsection{Interpretations of the Crowd Knowledge}
\label{sec:crowdknowledge}
We represent the interpretation of the answers provided by the crowd in three knowledge bases modeled as fuzzy sets: $KB^+$, $KB^-$, and $KB^\sim$.
$KB^+$ comprises the triples that should belong to the data set, while those that should not exist according to the crowd compose $KB^-$; finally, $KB^\sim$ contains the associations that the crowd could not establish because of a lack of knowledge.
For example, a triple in $KB^+$ indicates that the crowd considers that this triple should be part of the RDF data set, e.g., {\sfs db:Madrid} is the capital of {\sfs db:Spain} as reported in Table~\ref{table:example_results1}. 
All the triples in these fuzzy sets are annotated with a membership degree $m$, which states how reliable  an answer from the crowd is. 
We have empirically observed that in some cases workers declare to be unfamiliar when evaluating some triple patterns, e.g., the country of {\sfs db:Chi\c{s}in\u{a}u}, although the platform reported high confidence on this answer (see Table~\ref{table:example_results1}).  
Therefore, in this work, we computed $m$ as the
average of the workers' confidence and normalized familiarity.

\begin{definition}{\it (Interpretations of the Crowd Knowledge).}
Given $D$ an RDF data set and {\it CROWD} a pool of human resources.
Let $D^*$ be a virtual data set such that it is composed of all the triples that {\it `should'} be in $D$.
The interpretations of the knowledge  of {\it CROWD} is defined as a 3-tuple:
$$KB = (KB^+, KB^-, KB^\sim)$$
where $KB^+, KB^-, KB^\sim$ are fuzzy sets over RDF data composed of  4-tuples (quads) of the form $(s,p,o,m)$ such that:
\begin{itemize}
\item $m \in [0,1]$ is the membership degree of the RDF triple $(s,p,o)$ to the corresponding fuzzy set,
\item $(s,p,o,m) \in KB^+$ iff $o$ is a constant and according to {\it CROWD} $(s,p,o)$ belongs to the virtual data set $D^*$,
 \item $(s,p,o,m)\in KB^-$ iff $o$ is a variable and according to {\it CROWD} $(s,p,o)$ doesn't belong to the virtual data set $D^*$, and
\item $(s,p,o,m) \in KB^\sim$ iff $o$ is a variable or a constant, and according to {\it CROWD} the membership of $(s,p,o)$ to the virtual data set $D^*$ is unknown.
\end{itemize}
\label{def:confidence}
\end{definition}

Given an RDF resource $s$ and a predicate $p$, we are also interested in representing the completeness of $s$ with respect to $p$ in the knowledge base $KB^+$.

\begin{definition}{\it (Completeness of an RDF Resource with Respect to a Predicate in $KB$.)}
Given an RDF resource $s$ and a predicate $p$ occurring in the knowledge base $KB$. Let $C_1$, ..., $C_n$ be the classes in the RDF data set $D$ such that $(s, a, C_1) \in D, ... , (s, a, C_n) \in D$. The completeness of $s$ with respect to $p$ in $KB$ is defined as follows:
\begin{equation*}
\hbox{\it Comp}_{KB}(s|p) :=
\left\{
\begin{array}{lll}
 \frac{M_{KB}(s|p)}{AM_D(C'|p)}   & \mbox{if } AM_D(C'|p) \neq 0 \\
 1  & \mbox{otherwise}
\end{array}
\right.
\end{equation*}
\begin{itemize}
\item $M_{KB}(s|p)= |\{o | (s,p,o) \in KB^+ \lor (s,p,o) \in KB^- \lor (s,p,o) \in KB^\sim\}|$
\item $AM_D(C'|p) = max(AM_D(C_1|p), ... ,AM_D(C_n|p))$
\end{itemize}
\end{definition}

Consider that the movie {\sfs db:Tower\_Heist} also belongs to the class {\sfs http://dbpedia.org/ontology/Film}, and the aggregated multiplicity of this class with respect to  the predicate {\sfs db-prop:producer} is 5. Suppose  the crowd has declared that this movie is produced by {\sfs db:Brian\_Grazer} and this fact is part of $KB^+$. Then, the completeness of   {\sfs db:Tower\_Heist} in  $KB^+$ is 0.2, indicating that 80\% of the producers of this movie have not been collected from the crowd.

%
%


\subsubsection{Measuring Disagreement}
\label{sec:disagreement}
Consider an RDF resource $s$ and predicate $p$.
The {\it CROWD} disagreement about the (non-)existence of RDF triples with subject $s$ and predicate $p$ is defined as follows:
\begin{equation}
D(s|p) = 1 - |m^+ - m^-|
\label{eq:disagreement}
\end{equation}
\begin{itemize}
\item $m^+ = average(\{m | (s,p,o,m) \in KB^+\})$
\item $m^- = average(\{m | (s,p,o,m) \in KB^-\})$
\end{itemize}
Disagreement values close to 0.0 indicate high consensus about the (non-)existence of a triple in the virtual data set $D^*$. To illustrate this, suppose that  {\it CROWD} is enquired to provide the {\sfs db-prop:producer} for the movie {\sfs db:Tower\_Heist} from the data set in Figure~\ref{fig:cms}, and the crowdsourced answers are:
\begin{inparaenum}[\itshape i\upshape)]
\item {\it ``Brian Grazer is a producer of Tower Heist''}, with a  membership degree of 0.90, i.e., $t1 \in KB^+$ with $t1=${\sfs (db:Lower\_Heist, db-prop:producer, db:Brian\_Grazer, 0.90)}; and 
\item {\it ``Tower Heist has no producers''}, with a membership degree of 0.05, i.e., $t2 \in KB^-$ with $t2=${\sfs (db:Lower\_Heist, db-prop:producer, \_:o, 0.05)}.\footnote{In RDF, existential variables are represented as blank nodes, denoted in this example as {\sfs \_:o}.} Suppose that $t1$ and $t2$ are the only triples in $KB^+$ and $KB^-$, respectively. The disagreement in {\it CROWD} about the producer of this movie is $1 - |0.90 - 0.05| = 0.15$.
\end{inparaenum}
Low disagreement suggests that {\it CROWD} confirms the (non-)existence of a certain fact.

\subsubsection{Measuring Uncertainty}
\label{sec:uncertainty}
Consider an RDF resource $s$ and predicate $p$.  The uncertainty of {\it CROWD} about the (non-)existence of RDF triples with subject $s$ and predicate $p$ is defined as follows:
\begin{equation}
U(s|p) = m^\sim \\
\hbox{, where $m^\sim = avg(\{m | (s,p,o,m) \in KB^\sim\})$}
\end{equation}
Uncertainty values close to 1.0 indicate that {\it CROWD} has shown to be unknowledgeable about the fact to be vetted. To illustrate this, suppose that when {\it CROWD} is inquired about providing the producers for the movie {\sfs db:Non-Stop\_(film)}, the obtained answer is $t \in KB^\sim$ with  $t=${\sfs (db:Non-Stop\_(film), db-prop:producer, \_:o1, 0.97)}.
High uncertainty values indicate that {\it CROWD} doesn't have the knowledge to answer this question, and hence it's not useful to further insist on an assessment of this fact from the crowd.

%

%% file: sections/approach-microtask.tex
\subsection{Microtask Manager}
\label{sec:microtask_manager}
The microtask manager creates the human tasks and submits them to the crowdsourcing platform. 
This component receives the triple patterns to be crowdsourced, with bindings produced during query execution. 
Consider our running example from Listing~\ref{motivatingquery1}: the bindings for {\sfs ?city} obtained when evaluating {\sfs (?city, a, dbpedia-yago:CapitalsInEurope)} are then used to create the microtasks. 
For each instance, the microtask manager exploits the semantics of the RDF resources to build rich user interfaces that facilitate the worker's task. For example, an RDF-Hunter human task displays {\sfs "Madrid"}  instead of {\sfs \scriptsize http://dbpedia.org/resource/Madrid}. 
Providing details like these in microtasks has proven to assist the crowd in effectively   providing the right answer~\cite{acosta2013}. 

In addition, the RDF-Hunter human tasks contain two types of questions. 
The first one is related to the existence of a value for the crowdsourced triple pattern. 
For example, for the triple pattern $t=${\sfs (db:Madrid, dbpedia-owl:country, ?country)} the task displays: {\it ``Does Madrid have a country?"}. 
We devise three possible answers here, such that $t$ can be directly mapped into the crowd knowledge bases: {\it ``Yes"} $\leadsto$ $t \in KB^+$, {\it ``No"} $\leadsto$ $t \in KB^-$, and {\it ``Not sure"} $\leadsto$ $t \in KB^\sim$. 
When the answer is {\it ``Yes"}, a second question requires the crowd to provide a specific value, for example,  {\it ``What is the country of Madrid?"}. 
The microtask manager aggregates the outcome of the tasks and stores them as RDF triples annotated with the corresponding  membership degree ($m$).

%% file: sections/approach-optimizer.tex
\subsection{Query Decomposer \&  SPARQL Query Engine}
\label{sec:decomposer}
The RDF-Hunter query decomposer automatically identifies those parts of a query $Q$ that will be handled through human computation ($S_{\it CROWD}$) and those executed against the RDF data set ($S_D$). 
The decomposer performs three main steps: 
 \begin{inparaenum}[\itshape 1\upshape)]
\item  partitioning of $Q$ into $S_D$ and $S_{\it CROWD}$;
\item generation of the set $SQ_D$ of sub-queries to be executed against the RDF data set; and
\item  generation of the set $SQ_{\it CROWD}$ of sub-queries to be posed against the crowd.
\end{inparaenum}
The decomposer proceeds as follows: triple patterns with constants in the predicate and object positions are added to $S_D$, while those with variables in the subject and object positions are added to $S_{\it CROWD}$. 
For example,  in the SPARQL query from Figure~\ref{fig:execution}, triple patterns {\it t1}, {\it t3}, and {\it t4} are inserted in $S_D$, and  {\it t2} is added to $S_{CROWD}$. Next, the triple patterns in $S_D$ are partitioned into sub-queries in a way that all the triple patterns that share the same subject variable are assigned to the same sub-query.   Similarly, triple patterns in $S_{\it CROWD}$ are grouped into sub-queries.
In our running example,  triple patterns {\it t1}, {\it t3}, and {\it t4} share the variable {\sfs ?movie}, hence they compose one sub-query in $SQ_D$; further, {\it t2} belongs to the only sub-query in $SQ_{\it CROWD}$.

%% file: sections/approach-engine.tex
\begin{figure}[t!]
\begin{center}
\includegraphics[width=0.90\textwidth]{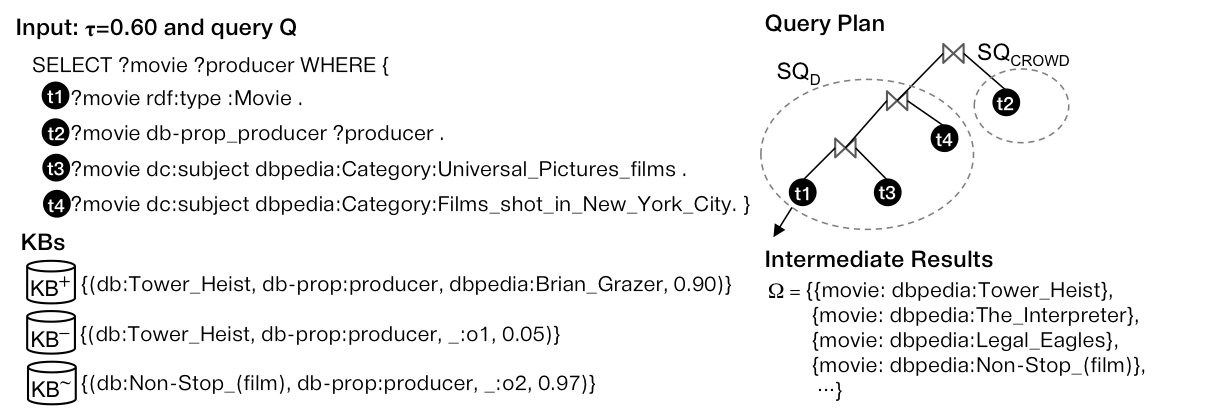}
\caption{Example of the execution of the RDF-Hunter decomposer and query engine}
\label{fig:execution}
\end{center}
\end{figure}

The RDF-Hunter  query engine executes SPARQL queries both against RDF data sets and the crowd to augment answer completeness. We propose an efficient algorithm (Algorithm~\ref{alg:engine}) that receives the decomposition of the query triple patterns into the sets $SQ_D$ and $SQ_{\it CROWD}$, and a threshold $\tau$ and outputs the answer of the query.
The algorithm generates left-linear plans where the most selective sub-queries are executed first, the number of joins is maximized (lines 1, 5, and 8), and Cartesian products are executed at the end (line 11).   Thus, intermediate results $\Omega$ and the number of human tasks are minimized. Figure~\ref{fig:execution} illustrates a query plan where the sub-query ({\it t1}, {\it t3}, and {\it t4}) against the RDF data set $D$ is executed first; the intermediate results $\Omega$ of this sub-query are used to instantiate the triple pattern {\it t2} that will be crowdsourced.
\begin{algorithm}[t]
\begin{scriptsize}
\caption{RDF-Hunter Query Execution Algorithm}
\label{alg:engine}
\KwIn{RDF data set $D$, a descomposition $(SQ_D, SQ_{CROWD})$,  $KB=(KB^+,KB^-,KB^\sim)$, and $\tau$.} 
\KwOut{ The query answer $\Omega$.  } 
\DontPrintSemicolon
 {$plan \gets sq$} // Where $sq$ is a sub-query from $SQ_D$ with the highest selectivity\\
 {$SQ_D' \gets SQ_D$}\\
 {$SQ_D \gets SQ_D - \{sq\}$}\\
// (1) Plan generation\\
\While{$SQ_D \cup SQ_{CROWD} \neq \emptyset$} {
    {Select $sq'$ from $SQ_{CROWD}$ such that $sq'$ shares a variable with $plan$}\\
    {$SQ_{CROWD} \gets SQ_{CROWD} - \{sq'\}$}\\
    {$plan \gets plan \cup \{sq'\}$}\\
    {Select from $SQ_{D}$ the sub-query $sq$ with the highest selectivity and that shares a variable with $plan$}\\
    {$SQ_{D} \gets SQ_{D} - \{sq\}$}\\
    {$plan \gets plan \cup \{sq\}$}\\
}
{$plan \gets plan \cup SQ_{CROWD} \cup  SQ_{D}$}\\
// (2) Execution of the plan\\
{ $ \Omega \gets \emptyset$ // Intermediate results}\\
\For{$pi \in plan$}{
    \If{$pi \in SQ_D'$}{
         {$pi' \gets instantiate(pi, \Omega)$ }\\
        {$\Omega \gets \Omega \Join [[pi']]_D$}\\
    }\Else{
        {$\Omega' \gets \Pi_{var(pi) \cap var(\Omega)} \Omega$}\\
        \For{$t=(s,p,o) \in pi$}{
               \For{$\mu \in \Omega'$} {
               \If{{\it Comp$_D$}$(\mu(s)|p)$ + {\it Comp$_{KB^+}$}$(\mu(s)|p) < 1.0$}{
                 \If{{\it P($\mu$(s),p) $> \tau$}}{
                     {Invoke the Microtask Manager with $(\mu(s), p, o)$}\\
		  }
               } 
		{$\Omega \gets \Omega \Join ([[(\mu(s), p, o)]]_D \cup [[(\mu(s), p, o)]]_{KB^+})$}\\
        }
%
        }
   } 
} 
\Return $\Omega$
\end{scriptsize}
\end{algorithm}
%
%
%

Given a triple pattern $t$ in a sub-query $pi$ assigned to the crowd, the algorithm checks if the instantiation of $t$ with the mappings from $\Omega$ can be evaluated against $D$ and $KB^+$, and produce complete results (line 22). If $D$ and $KB^+$ are not complete (lines 23-24), the algorithm verifies if the crowd could potentially collect a complete answer, i.e.,
it checks if $P(\mu(s),p)>\tau$ holds. The  threshold $\tau$  is provided by the user, and $P(\mu(s),p)$ corresponds to the probability of crowdsourcing
the evaluation of $t$ where $\mu$(s) and $p$ are in the subject and predicate positions of $t$, respectively:
\begin{equation}
P(\mu(s),p) = \alpha \cdot (1 -  Comp(s|p)) + (1-\alpha) \cdot T(D(s|p), U(s|p))
\label{eq:com}
\end{equation}
%
\begin{itemize}
\item $\alpha \in [0.0,1.0]$ is a score to weight the importance of the data set completeness and the crowd knowledge,
\item $Comp(s|p)$ is defined as $Comp_D(s|p) + {\it Comp}_{KB^+}(s|p)$,
\item $D(s|p)$ and $U(s|p)$ correspond to the disagreement and uncertainty levels,
\item $T$ is a T-norm function to combine the values of disagreement and uncertainty.
We compute $T$ as the G\"{o}del T-norm, also called Minimum T-norm, which represents a weak conjunction of fuzzy sets. Since the system aims at crowdsourcing triples with high levels of disagreement but low uncertainty, we have applied the G\"{o}del T-norm as follows:
$T(D(s|p), U(s|p)) = \min\{D(s|p), 1-U(s|p)\}$
\end{itemize}
If $P(\mu(s),p)$~$> \tau$ holds, the algorithm invokes the microtask manager which creates the corresponding microtasks and submits them to the crowdsourcing platform.
The algorithm terminates when all the sub-queries are evaluated and produces  $\Omega$.

We illustrate the behavior of Algorithm~\ref{alg:engine} (lines 21-25) when $AM_D$({\sfs schema.org:Movie}$|$ {\sfs db-prop:producer})=3, the plan  and intermediate results $\Omega$ from Figure~\ref{fig:execution} are considered.

{\bf Iteration 1:} The algorithm selects the first element of the intermediate results $\Omega$, 
$\mu${\sfs (movie) = 
db:Tower\_Heist}. Given that  $M_D${\sfs (db:Tower\_Heist}$|${\sfs db-prop:producer) = 0} (see Figure~\ref{fig:cms}) and  $M_{KB^+}${\sfs (db:Tower\_Heist}$|${\sfs db-prop:producer) = 2} the completeness of $\mu(s)$ w.r.t. $p$ is $0.33$. The algorithm then computes the probability of evaluating the triple pattern {\sfs (db:Tower\_Heist, db-prop:producer, ?producer)} against the crowd (line 23).  The crowd knowledge bases $KB^+$ and $KB^-$ have information about this triple pattern, and applying Equation~\ref{eq:disagreement}, the algorithm obtains that $D(\mu(s)|p)=0.15$. Notice that this triple pattern is not in $KB^\sim$, hence $U(\mu(s)|p)=0$. The result of applying Equation~\ref{eq:com} is $P(\mu(s),p)=0.5 \cdot (1-(0+0.33)) + 0.5 \cdot min(\{0.15, 1-0\}) = 0.41$. This value is lower than $\tau=0.60$, then this pattern is not submitted to the crowd.

{\bf Iteration 2:} The next instance is $\mu${\sfs (movie) = db:The\_Interpreter}. According to Figure~\ref{fig:cms}, $M_D${\sfs (db:The\_Interpreter}$|${\sfs db-prop:producer) = 3}, then the completeness of $\mu(s)$ w.r.t. $p$ in the data set is $1.0$. Since the values of $p$ for $\mu(s)$ are complete, according to the algorithm on line 22,  this triple pattern is not crowdsourced.

{\bf Iteration 3:} The algorithm processes  $\mu${\sfs (movie) = db:Legal\_Eagles}, with $M_D${\sfs (db:Legal\_Eagles}$| ${\sfs db-prop:producer) = 2} (see Figure~\ref{fig:cms}). The completeness of $\mu(s)$ w.r.t. $p$ is $0.667$, which suggests that this triple pattern might be subject to crowdsourcing (line 22).  The crowd knowledge bases don't have information about this triple pattern, therefore $m^+=0$, $m^-=0$, $m^\sim=0$; the disagreement and uncertainty are $D(\mu(s)|p)=1.0$ and $U(\mu(s)|p)=0.0$, respectively. Applying Equation~\ref{eq:com},  the algorithm obtains that $P(\mu(s),p)=0.5 \cdot (1-(0.667+0)) + 0.5 \cdot min(\{1.0, 1-0.0\}) = 0.667$ (line 23). This value is higher than $\tau=0.60$, in consequence, this pattern is submitted to the crowd.

{\bf Iteration 4:} The next instance is $\mu${\sfs (movie) = db:Non-Stop\_(film)}, with $M_D${\sfs (db:Non-Stop\_(film)}$|${\sfs db-prop:producer) = 0} (see Figure~\ref{fig:cms}), hence the algorithm checks whether this triple pattern is crowdsourced (line 22). $KB^+$ and $KB^-$ don't have information about this triple pattern, therefore, $D(\mu(s)|p)=1.0$.  However, $KB^\sim$ states that this triple pattern has uncertainty $U(\mu(s)|p)=0.97$. The algorithm computes:  $P(\mu(s),p)=0.5 \cdot (1-(0+0)) + 0.5 \cdot min(\{1, 1-0.97\}) = 0.515$, therefore, this pattern is not crowdsourced.

The algorithm then joins the intermediate results $\Omega$ with the corresponding instances from the data set $D$ and the crowd knowledge $KB^+$ (line 25). Considering the RDF data set from Figure~\ref{fig:cms}, the evaluation of the running query with traditional SPARQL engines yields only $5$ results: 3 producers for {\sfs db:The\_Interpreter} and 2 producers for {\sfs db:Legal\_Eagles}. With RDF-Hunter, the query answer in addition contains the producer for the movies {\sfs db:Tower\_Heist}, and the potential answers of evaluating against the crowd the triple pattern {\sfs (db:Legal\_Eagles, db-prop:producer, ?producer)} as identified in Iteration 3.


%% file: sections/experiments.tex
\section{Experimental Study}
\label{sec:experiments}
\vspace{-0.15cm}
We conducted an empirical evaluation to assess the effectiveness of RDF-Hunter to augment the answer completeness of SPARQL queries via microtasks.  Below we describe the configuration settings used in our experiments.  
\input{sections/experimentalsetting}

We executed the benchmark queries with RDF-Hunter and crowdsourced a total of 502 RDF triples. We collected 1,619 answers from the crowd (see Table~\ref{table:res}). 


\begin{table}[t!]
\centering
\caption{Results when executing the benchmark with RDF-Hunter}
\begin{tabular}{|l|c|c|c|c|}
\hline
{\bf Knowledge} & {\bf Result Set} & {\bf \# Crowdsourced} & {\bf \# Total Crowd} & {\bf F-Measure}\\
{\bf Domain} & {\bf Completeness} & {\bf RDF Triples} & {\bf Responses} & \\
& {\scriptsize \bf (min; max)} &  {\bf \scriptsize (w/o Gold Units)} &{\bf \scriptsize (w/o Gold Unit Responses)}  & \\
\hline
Life Sciences & (0.03; 0.92) & 82 & 250 & 0.96\\
History& (0.04; 0.91) & 160 & 476 & 0.89\\
Music& (0.36; 0.80) & 71& 204 & 0.84\\
Sports& (0.25; 0.86) & 69 & 199 & 0.91\\
Movies& (0.46; 0.89) & 120& 490 & 0.95\\
\hline
Total & - & {\bf 502} & {\bf 1,619} & - \\
\hline
\end{tabular}\\
\label{table:res}
\end{table}

\subsection{Results: Accuracy of Crowd Answers}
\label{sec:rescompletness}
\begin{figure}[h!]
\centering
\subfigure[Heat map of recall]{\includegraphics[width=0.48\linewidth]{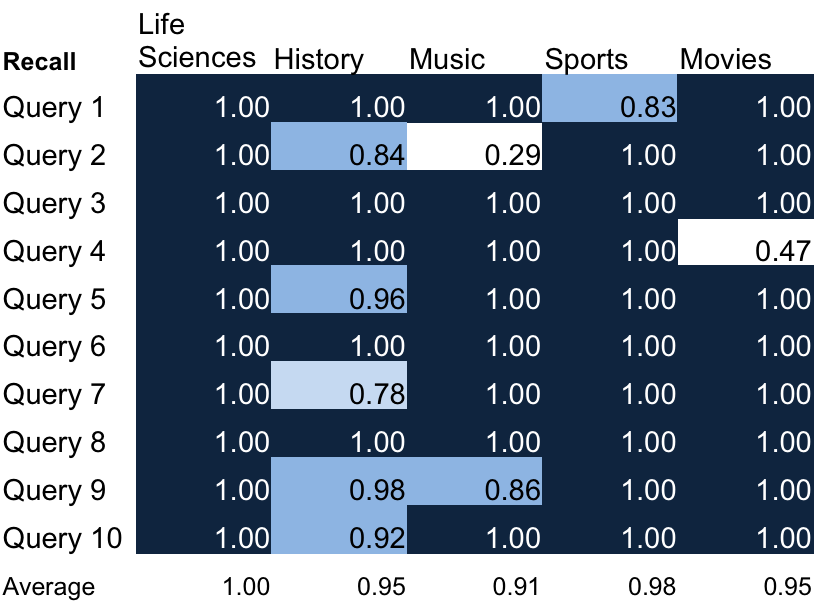}\label{fig:recall}} \hbox{ }
\subfigure[Heat map of precision]{\includegraphics[width=0.48\linewidth]{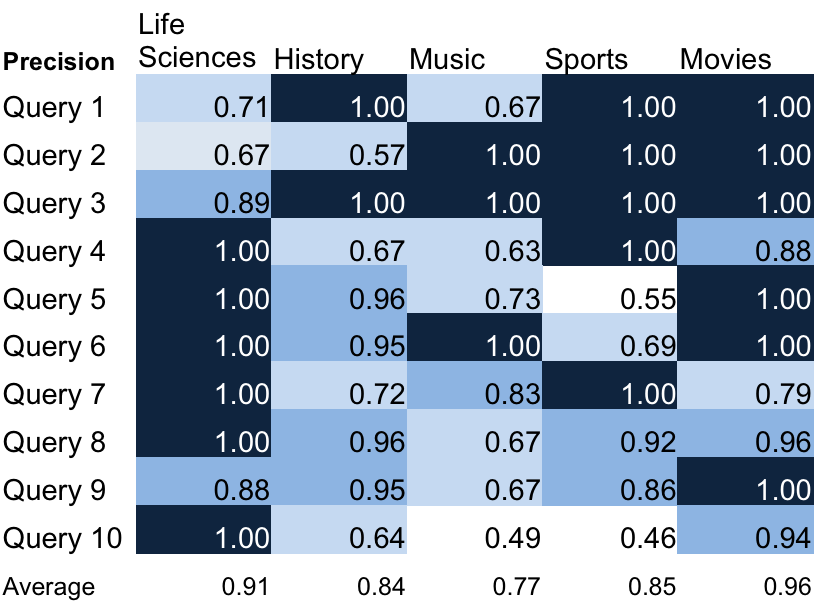}\label{fig:precision}} \hbox{ }
\caption{Heat maps of recall and precision achieved with RDF-Hunter}
\label{fig:heatmap}
\end{figure}

We report on results of precision and recall using heat maps (see Figure~\ref{fig:heatmap}). The darkest color represents values of precision or recall equal to 1.0. Columns correspond to the five knowledge domains, while rows represent the benchmark queries. 
Altogether, the crowd was able to respond to 21 out of 50 queries with both precision and recall equal to 1.0, and the crowd achieved accuracy values ranging from 0.84 to 0.96 (see Table~\ref{table:res}).

Figure~\ref{fig:recall} reports on the values of recall. We can observe that in 41 out of 50 queries, the crowd was able to answer all the missing values. Furthermore, for 48 queries the achieved recall is greater than 0.77. The only two cases where the crowd achieved low recall are in queries {\sfs Music Query 2} and {\sfs Movies Query 4}. 
The questions for these queries were: ``What is the name of {\it an American blues musician}?''  and ``What is the gross of {\it a movie}?'', respectively.  
Although the crowd showed to be skilled in these two domains in general, there  are  predicates within these domains where the crowd doesn't exhibit high levels of expertise. 
This observation provides evidence of the importance of the RDF-Hunter triple-based approach on the identification of portions of the domain where the crowd is knowledgeable. Thus, in subsequent requests, RDF-Hunter will exploit this knowledge to avoid crowdsourcing these two questions again. 

The values of precision are reported on Figure~\ref{fig:precision}. The crowd was able to provide correct answers for 22 out of 50 queries. Furthermore, the crowd achieved a precision greater than 0.75 in 35 queries.  
The heat map clearly shows the heterogeneity of the level of crowd expertise, and similarly to recall, precision values support the importance of the RDF-Hunter triple-based approach. 
\vspace{-0.5cm}

\subsection{Results: Crowd's Confidence \& Familiarity}
The values of crowd's confidence  and familiarity are used by RDF-Hunter to annotate the triples retrieved from the crowd. 
These annotations represent the membership degree ($m$) of the triples to each of the crowd knowledge bases, and guide the RDF-Hunter execution algorithm to devise an effective query evaluation.
RDF-Hunter captures the crowd's confidence provided by CrowdFlower as {\it worker trust}, which suggests how reliable the answer provided by a worker is.
In addition, the crowd's familiarity with the topic is collected via required questions embedded in each microtask. 

The average and standard deviation values of the crowd's confidence obtained from CrowdFlower are: (Life Sciences, $0.92 \pm 0.07$), (History, $0.93 \pm 0.07$), (Music, $0.95 \pm 0.07$), (Sports, $0.94 \pm 0.07$), (Movies, $0.94 \pm 0.07$). 
The crowd's confidence on average is very high indicating that the majority of the crowd answers are reliable. 
However, the homogeneity of these results suggests that considering only the crowd's confidence is not enough to model the membership degree $m$ of the crowd answers. 

\begin{figure*}[b!]
\begin{center}
\includegraphics[width=0.32\textwidth]{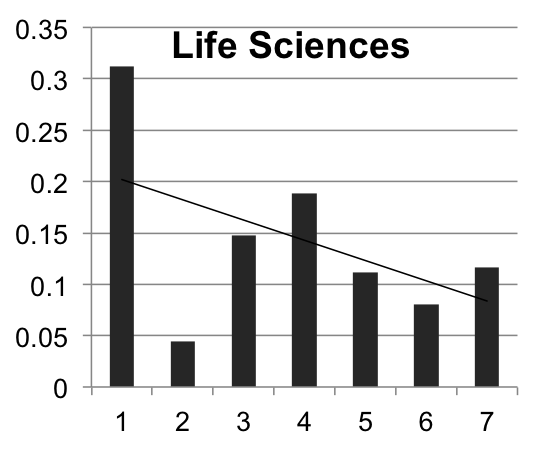}
\includegraphics[width=0.32\textwidth]{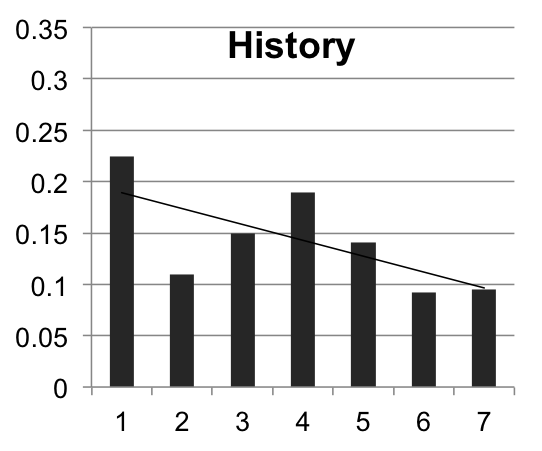}
\includegraphics[width=0.32\textwidth]{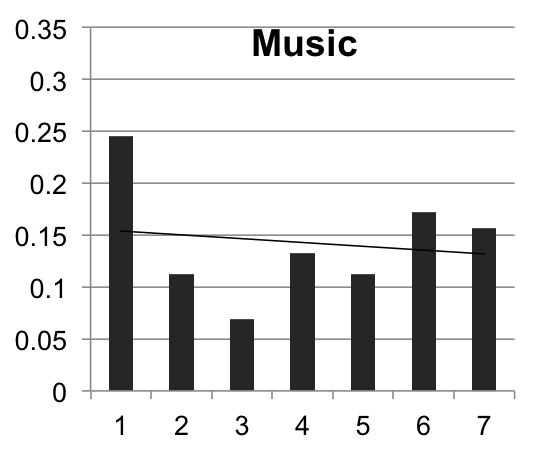}\\
\includegraphics[width=0.32\textwidth]{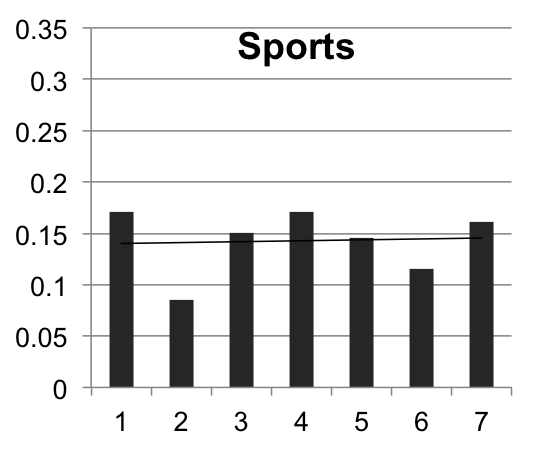}
\includegraphics[width=0.32\textwidth]{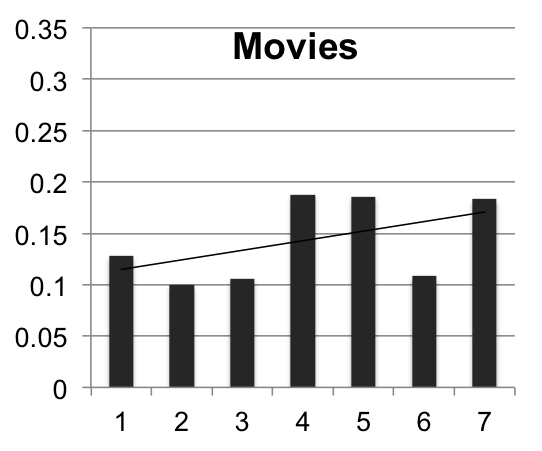}
\caption{Histogram and trend line of the crowd's familiarity with tasks from five different knowledge domains: Life Sciences, History, Sports, Movies, and Music. The $x$ axis represents the familiarity score from 1 to 7; the $y$ is the proportion of workers}
\label{fig:familiarity}
\end{center}
\end{figure*}

Figure~\ref{fig:familiarity} reports the histograms for familiarity scores in a scale from 1 to 7. 
The familiarity values provide insights on how difficult the workers perceive a given task. In our experiments, according to the reported familiarity, the topics with higher familiarity were Sports and Movies. The triples retrieved from the crowd in these topics are annotated with high values of $m$. 
On the other hand, the more challenging domain was Life Sciences, indicating that some crowdsourced answers in this domain are annotated with lower values of $m$ in comparison to Sports or Movies. 
Thus, values of familiarity combined with crowd's confidence allow for more faithful values of membership $m$.

%% file: sections/experimentalsetting.tex

\noindent
{\bf Query Benchmark:}
We created an extensive benchmark of 50 queries\footnote{Queries are available at \url{http://people.aifb.kit.edu/mac/rdf-hunter}} by analyzing sub-queries answerable for the DBpedia SPARQL endpoint; we designed queries that yield incomplete results,   varying the percentage of result set completeness from 0.03 to 0.92 (see Table~\ref{table:res}).  To test the knowledge of the crowd, we crafted 10 queries about different topics in five domains: {\it Life Sciences}, {\it History}, {\it Music}, {\it Sports}, and {\it Movies}. Queries are composed of basic graph patterns of between three and six triple patterns.

\noindent{\bf Gold Standard:} We created a gold standard of the form ({\it triple pattern}, {\it answers}). 
For each {\it triple pattern} in the benchmark queries, we retrieved the {\it answers} produced by the endpoint. When RDF-Hunter decides to submit a triple pattern $t$ to the crowd, a {\it triple pattern} from the gold standard with the same predicate of $t$ is crowdsourced. The answers from the crowd are then compared to {\it answers} to compute accuracy.


\noindent
{\bf Evaluation Metrics:}
\begin{inparaenum}[\itshape i\upshape)]
 \item {\it Precision} $(P)$: Given a query $q$, precision measures the fraction of the answers collected from the crowd during the hybrid evaluation of $q$  that actually correspond to answers of $q$; values of precision close to 1.0 show that the crowd outputs correct answers for $q$.  
 \item {\it Recall} $(R)$: Given a query $q$, recall measures the fraction of the missing answers of $q$ that are collected from the crowd during the hybrid evaluation of $q$; values of recall equal to 1.0 indicate that the crowd is able to produce all of the missing  answers of $q$. 
\item {\it F-Measure}: Combines the values of precision and recall to measure the accuracy of the crowd output; the F-Measure is computed as follows: $2\cdot \frac{P \cdot R}{P+R}$.
  \end{inparaenum}
  
\noindent
{\bf Implementation:} RDF-Hunter is implemented in Python 2.7.6. and CrowdFlower is  used as the crowdsourcing platform. We set up RDF-Hunter with $\tau=0.02, \alpha=0.5$. The crowd knowledge bases $KB^+$, $KB^-$, and $KB^\sim$ were initially empty.

\noindent
{\bf Crowdsourcing Configuration:}
\begin{inparaenum}[\itshape i\upshape)]
 \item {\it Task granularity}: In each task,  we asked the workers to solve a maximum of four  different questions; 
each question corresponds to one RDF triple. 
\item 
{\it Payments}: The monetary reward was fixed to 0.07 US dollars per task, i.e., we paid 0.0175 US dollar per RDF triple.
\item
{\it Judgments}: We configured the CrowdFlower platform to collect at least three answers for each question.
\item {\it Gold Units}:  Correspond to verification questions used by CrowdFlower to filter low-quality workers. In this work, the gold units were generated from the gold standard. The gold unit distribution was set to 10:90, i.e., for each 100 triples in the gold standard, 10 were gold units. 
 \end{inparaenum}

%% file: sections/conclusions.tex
\section{Conclusions and Outlook}
\label{sec:conclusions}
We defined RDF-Hunter, the first hybrid engine for SPARQL query answering with human computation. 
RDF-Hunter supports  crowdsourcing to enhance  the completeness of Linked Data sets at query processing time. 
Both the quality model and the novel query engine enable RDF-Hunter to  automatically  decide which parts of a SPARQL query should resort to human computation, according to the data set quality. 
We designed an extensive query benchmark of 50 queries over the DBpedia data set. 
Empirical results confirm that hybrid human/computer systems can effectively increase the completeness of SPARQL query answers. RDF-Hunter achieved F-measure values ranging from 0.84 to 0.96. 
In the future, we will focus on optimizations such us batching~\cite{qurk2011} and prioritization~\cite{deco2012technicalreport} of human computation tasks to provide a  more efficient, adaptive, and scalable RDF-Hunter query engine. 


%% file: rdf-hunter.bbl
\begin{thebibliography}{10}

\bibitem{acosta2013}
M.~Acosta, A.~Zaveri, E.~Simperl, D.~Kontokostas, S.~Auer, and J.~Lehmann.
\newblock Crowdsourcing linked data quality assessment.
\newblock In {\em 12th International Semantic Web Conference, Sydney, NSW,
  Australia, October 21-25, 2013, Proceedings, Part {II}}, pages 260--276,
  2013.

\bibitem{DBLP:conf/sigmod/AmsterdamerDMNS14}
Y.~Amsterdamer, S.~B. Davidson, T.~Milo, S.~Novgorodov, and A.~Somech.
\newblock {OASSIS:} query driven crowd mining.
\newblock In {\em International Conference on Management of Data, {SIGMOD}
  2014, Snowbird, UT, USA, June 22-27, 2014}, pages 589--600, 2014.

\bibitem{franklin2011}
M.~Franklin, D.~Kossmann, T.~Kraska, S.~Ramesh, and R.~Xin.
\newblock {CrowdDB: answering queries with crowdsourcing}.
\newblock In {\em Proceedings of the 2011 ACM SIGMOD International Conference
  on Management of Data}, pages 61--72, 2011.

\bibitem{DBLP:journals/pvldb/0002KMMO12}
A.~Marcus, D.~R. Karger, S.~Madden, R.~Miller, and S.~Oh.
\newblock Counting with the crowd.
\newblock {\em {PVLDB}}, 6(2):109--120, 2012.

\bibitem{qurk2011}
A.~Marcus, E.~Wu, D.~Karger, S.~Madden, and R.~Miller.
\newblock Human-powered sorts and joins.
\newblock {\em Proceedings of VLDB Endowment}, 5(1):13--24, 2011.

\bibitem{deco2012vldb}
H.~Park, R.~Pang, A.~G. Parameswaran, H.~Garcia-Molina, N.~Polyzotis, and
  J.~Widom.
\newblock {Deco: A System for Declarative Crowdsourcing}.
\newblock {\em Proceedings of VLDB Endowment}, 5(12):1990--1993, 2012.

\bibitem{deco2012technicalreport}
H.~Park, A.~Parameswaran, and J.~Widom.
\newblock {Query Processing over Crowdsourced Data}.
\newblock Technical report, Stanford University, September 2012.

\bibitem{DBLP:journals/pvldb/ParkW13}
H.~Park and J.~Widom.
\newblock Query optimization over crowdsourced data.
\newblock {\em {PVLDB}}, 6(10):781--792, 2013.

\bibitem{LOD2014}
M.~Schmachtenberg, C.~Bizer, and H.~Paulheim.
\newblock Adoption of the linked data best practices in different topical
  domains.
\newblock In {\em The Semantic Web -- ISWC 2014}, volume 8796 of {\em Lecture
  Notes in Computer Science}, pages 245--260. Springer International
  Publishing, 2014.

\bibitem{DBLP:conf/icde/TrushkowskyKFS13}
B.~Trushkowsky, T.~Kraska, M.~J. Franklin, and P.~Sarkar.
\newblock Crowdsourced enumeration queries.
\newblock In {\em 29th {IEEE} International Conference on Data Engineering,
  {ICDE} 2013, Brisbane, Australia, April 8-12, 2013}, pages 673--684, 2013.

\end{thebibliography}
